# An LMP O(log n)-Approximation Algorithm for Node Weighted Prize Collecting Steiner Tree

Jochen Könemann, Sina Sadeghian, Laura Sanità. *

**Abstract**

In the *node-weighted prize-collecting Steiner tree problem* (NW-PCST) we are given an undirected graph $G = (V, E)$, non-negative costs $c(v)$ and penalties $\pi(v)$ for each $v \in V$. The goal is to find a tree $T$ that minimizes the total cost of the vertices spanned by $T$ plus the total penalty of vertices not in $T$. This problem is well-known to be *set-cover* hard to approximate. Moss and Rabani (STOC'01) presented a primal-dual Lagrangean-multiplier-preserving $O(\ln |V|)$-approximation algorithm for this problem. We show a serious problem with the algorithm, and present a new, fundamentally different primal-dual method achieving the same performance guarantee. Our algorithm introduces several novel features to the primal-dual method that may be of independent interest.

## 1 Introduction

The *node-weighted Steiner tree* problem (NWST) is a fundamental and well understood network design problem, where we are given an $n$-node, undirected graph $G = (V, E)$, a non-negative cost $c(v)$ for each vertex $v \in V$, and a set of *terminals* $R \subseteq V$. The goal is then to find a tree $T \supseteq R$ that has minimum total cost $\sum_{v \in T} c(v)$.

In this paper, we consider the *prize-collecting* version of the problem. As in standard NWST, the input in an instance of the *node-weighted prize-collecting Steiner tree* problem (NW-PCST) again consists of an $n$-node, undirected graph $G = (V, E)$ and a non-negative cost $c(v)$ for each vertex $v \in V$, but instead of a set of terminals we have now a non-negative *penalty* value $\pi(v)$ for each node $v \in V$. We wish to find a tree $T$ that minimizes

$$\sum_{v \in T} c(v) + \sum_{v \in V \setminus T} \pi(v).$$

Both NWST and NW-PCST have numerous practical and theoretical applications (e.g., [5, 9, 16]), and are well-known to be NP-hard. From an approximation point of view, there is a relatively straight-forward, approximation-factor preserving reduction from the *set-cover* problem, and therefore no $o(\ln n)$-approximation algorithm exists for either one, unless NP $\subseteq$ DTIME($n^{\text{polylog}(n)}$) [6, 11]. In fact, NWST and NW-PCST are significantly harder than the corresponding edge-weighted variants, for which constant approximation algorithms are known [7, 10, 15, 2, 3, 1].

---

*Department of Combinatorics and Optimization, University of Waterloo, emails: {jochen,s3sadegh,lsanita}@uwaterloo.ca

Klein and Ravi [11] showed that NWST admits an $O(\ln n)$ approximation, matching the above in-approximability bound. Guha et al. [9] later gave a primal-dual interpretation of this algorithm using a natural linear programming formulation of the problem.

The focus of this paper will be the approximability of the prize-collecting variant of the problem. We will in fact address the *rooted* version of the problem, where a specific root vertex $r \in V$ has to be part of the output tree $T$. Clearly, any algorithm for the rooted NW-PCST immediately yields an algorithm with the same performance guarantee for the unrooted version. The main result of this paper is stated in the following theorem.

**Theorem 1.1.** *There is a* Lagrangean multiplier preserving (LMP) $O(\ln n)$*-approximation algorithm for NW-PCST; i.e., there is an algorithm that, given an NW-PCST instance, computes a tree $T$ containing the root $r$ such that*

$$\sum_{v \in T} c(v) + \alpha \sum_{v \notin T} \pi(v) \leq \alpha \, opt,$$

*where opt is the value of an optimum solution to the problem, and $\alpha = O(\ln n)$.*

We note that it is reasonably straight forward to obtain a non-LMP $O(\ln n)$-approximation algorithm for NW-PCST via a standard *threshold rounding* approach for the natural LP formulation of the problem [1]. Ensuring the LMP property is highly non-trivial, however, and of crucial importance in the design of approximation algorithms for *partial* versions of NWST via the Lagrangean framework of Chudak et al. [4]. Two such partial problems that were considered in [14] are the *quota* and *budget* versions of NWST. In the former, we are given a non-negative profit $p(v)$ for each vertex $v \in V$ and a quota $Q > 0$, and we wish to find a connected set $T$ of smallest cost whose vertices have profit at least $Q$. In the latter problem, we are given a budget $B > 0$, and wish to find a connected set $T$ of largest total profit whose cost is at most $B$.

The correctness of Theorem 1.1 was previously claimed by Moss and Rabani [13, 14]; the primal-dual algorithm presented there does, however, have a crucial technical mistake that appears not to have a simple fix, as we explain in the next subsection. For this reason, we present a fundamentally different primal-dual approach. Beside proving correctness of Theorem 1.1, and therefore establishing correctness of results that rely on using Theorem 1.1 (e.g. approximation results for the previous mentioned partial NWST problems [14] or for problems arising in the study of contagion processes in networks [12, 8]), our algorithm features several new ideas that it adds to the known primal-dual repertoire, and therefore might be of independent interest.

We begin by presenting the natural integer linear programming formulation for NW-PCST (see also [14]). Afterwards, we give a high-level description of Moss and Rabani's algorithm, and show an example where it does not perform correctly. We then outline our algorithm, highlighting the main novelties with respect to currently known primal-dual approaches.

## 1.1 LP Formulation for NW-PCST

The integer program has a variable $x_v$ for each vertex $v \in V$ that has value 1 if $v$ is part of the output tree $T$, and $x_v = 0$ otherwise. We also have variables $z_S$ for all $S \subseteq V'$ where $V' = V \setminus \{r\}$.

We let $z_S = 1$ if $S$ is the set of vertices not spanned by $T$, and $z_S = 0$ otherwise. In the following we let $\Gamma(S)$ for a set $S \subseteq V$ be the set of all vertices $u \in V \setminus S$ that have a neighbour in $S$, and we let $\pi(S)$ be equal to $\sum_{v \in S} \pi(v)$.

$$
\begin{aligned}
\min \quad & \sum_{v \in V'} c(v) x_v + \sum_{S \subseteq V'} \pi(S) z_S & & \text{(P)} \\
\text{s.t.} \quad & \sum_{v \in \Gamma(S)} x_v + \sum_{U \mid S \subseteq U} z_U \geq 1 & & \forall S \subseteq V', \\
& x_v + \sum_{U \mid v \in U} z_U \geq 1 & & \forall v \in V', \\
& x_v \in \{0, 1\} & & \forall v \in V', \\
& z_S \in \{0, 1\} & & \forall S \subseteq V',
\end{aligned}
$$

We let (LP) be the linear programming relaxation of (P), obtained by replacing integrality constraints by non-negativity. Its LP dual is as follows.

$$
\begin{aligned}
\max \quad & \sum_{S \subseteq V'} y_S + \sum_{v \in V'} p_v & & \text{(D}^0\text{)} \\
\text{s.t.} \quad & \sum_{S \mid v \in \Gamma(S)} y_S + p_v \leq c(v) & & \forall v \in V' \\
& \sum_{U \subseteq S} y_U + \sum_{v \in S} p_v \leq \pi(S) & & \forall S \subseteq V' \\
& y \geq 0 \\
& p \geq 0
\end{aligned}
$$

Call a vertex $v \in V$ *cheap* if $c(v) \leq \pi(v)$, and *expensive* otherwise. Just like in [14], we obtain a *reduced* version of (D$^0$) by setting $p_v = c(v)$ whenever $v$ is a cheap vertex, and $p_v = \pi(v)$ otherwise. We then define the *reduced cost* $\bar{c}(v)$ of vertex $v$ to be 0 if $v$ is cheap, and we let it be $c(v) - \pi(v)$ otherwise. Similarly, we let the *reduced penalty* $\bar{\pi}(v)$ be $\pi(v) - c(v)$ if $v$ is cheap, and 0 otherwise. The reduced dual of (P) is then:

$$
\begin{aligned}
\max \quad & \sum_{S \subseteq V'} y_S + p(V) & & \text{(D)} \\
\text{s.t.} \quad & \sum_{S \mid v \in \Gamma(S)} y_S \leq \bar{c}(v) & & \forall v \in V' & & (1) \\
& \sum_{U \subseteq S} y_U \leq \sum_{v \in S} \bar{\pi}(v) & & \forall S \subseteq V' & & (2) \\
& y \geq 0
\end{aligned}
$$

## 1.2 Moss & Rabani's algorithm

The algorithm in [14] computes a dual solution for (D) using a *monotone* growing process. Initially, we let $\mathcal{C}$ be the set of all inclusion-wise maximal connected components of the graph induced by the cheap vertices and the root. All such components but the one containing the root are *active*.

The algorithm raises the dual variables $y_S$ corresponding to all active components in $\mathcal{C}$ uniformly, maintaining feasibility for (D). When constraint (2) becomes tight for some active component $C$, $C$ becomes inactive and its dual value is not increased anymore. When constraint (1) becomes tight for some expensive vertex $v$, all the (active and inactive) components adjacent to $v$ are removed from $\mathcal{C}$ and merged to form one new component $C'$ given by $C' := \{v\} \cup \{C \in \mathcal{C} : v \in \Gamma(C)\}$. This new component is active if it does not contain the root, and inactive otherwise. Then, the dual growing continues and the algorithm stops when no active component remains. At the end of the algorithm, the output is a tree $T$ contained in the connected component in $\mathcal{C}$ containing the root, and the dual solution $(y, p)$ for $(D^0)$. The claim is that

$$\sum_{v \in T} c(v) + \alpha \sum_{v \notin T} \pi(v) \leq \alpha \left( \sum_{S \subseteq V'} y_S + \sum_{v \in V'} p_v \right)$$

with $\alpha = O(\lg n)$.

We exhibit a counterexample where the above process finds a dual solution whose value is a factor of $\approx n$ lower than the cost of an optimum solution. The starting point for this example is the observation that the set-cover problem is a special case NWST, and that known LP-based $O(\ln n)$-analyses for set-cover use dual-fitting, or direct primal rounding approaches. So far, primal-dual algorithms based on a monotone dual growing process are only known to have performance ratio equal to the maximum *frequency* of any element (e.g., see [17]), but no better. Since Moss and Rabani's algorithm is indeed based on a monotone primal-dual process, this would (interestingly) lead to an algorithm of the same type for set-cover. However, as we now show, this is not the case.

Indeed, the instance shown in Figure 1 draws motivation from the usual set-cover reduction to NWST. The graph is obtained by taking a complete bipartite graph with cheap vertices $u_1, \ldots, u_n$ on one side and expensive vertices $v_1, \ldots, v_n$ on the other side. For each $1 \leq i \leq n$, another cheap vertex $w_i$ is attached to each expensive vertex $v_i$. All these cheap vertices have cost 0 and penalty $n$, while all these expensive vertices have penalty 0 and cost $n + 1$. Finally, we attach to $u_1$ an expensive vertex of cost 2 and penalty 0, that in turn is attached to the root.

When running the algorithm of [14] on this instance, at the beginning the cheap vertices $u_1, \ldots, u_n$ and $w_1, \ldots, w_n$ form singleton active components. In the first step, the dual values $y_{\{u_i\}}$ and $y_{\{w_i\}}$ ($i = 1, \ldots, n$) are increased by 1 and constraint (1) becomes tight for all expensive vertices $v_1, \ldots, v_n$. This basically implies that all such expensive vertices will join together to form a single component $C$ containing all vertices $u_i, v_i, w_i$, for $i = 1, \ldots, n$. The dual value $y_C$ of this new active component will then be increased by 1 where the expensive vertex adjacent to the root becomes tight as well. At this point the entire graph forms one inactive component, and the algorithm terminates. The total value of the dual solution $y$ computed by the algorithm is $O(n)$, while the optimal solution is any spanning tree of the whole graph, that has value $\Omega(n^2)$.

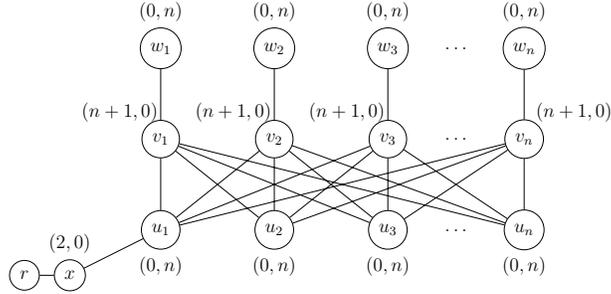

Figure 1: Counter example for the dual solution proposed in [13]. The pairs of numbers on vertices show the values of cost and penalty respectively.

## 1.3  The new algorithm

Our algorithm follows a dual-growing approach, as does that of Moss and Rabani. Unlike their algorithm, however, ours is not monotone! Instead, it is inspired by Guha et al.'s primal-dual view of Klein and Ravi's NWST algorithm. Like the algorithm in [9], our method works in phases. In each phase but the last we construct trees that merge several of the components from the previous phase. When the algorithm terminates, the cost of the tree connected to the root as well as the penalty of all vertices not spanned by this tree are approximately charged to a feasible solution to ($D^0$). Both our algorithm and its analysis depart from the standard primal-dual approach taken in [9] in several ways. We highlight the two most important new features.

First, a standard primal-dual strategy for node-weighted problems goes as follows: whenever at least two active components tighten constraint (1) of a vertex $v$, then a tree connecting these components and $v$ is built. Our algorithm instead considers connecting *active* and *inactive* components neighbouring $v$, but does so only if the total dual of participating components is *large* enough. We stress that the algorithm may build a tree even if only one of the participating components is active.

The second main difference of our methods lies in the charging argument that accounts for the cost of the final tree produced. For this, we identify a unique *core* for each set $S$ in the support of the dual solution produced. The cost of the computed final tree is charged to only those cores that it spans. We then combine this fact with a non-standard potential function based argument to show our claimed approximation guarantee.

Overall, both our algorithm and its analysis are significantly more involved than the ones of Moss and Rabani: in our opinion, this might be consistent with the fact that classical primal-dual approaches based on monotone growing seem to fail for NW-PCST, and therefore additional ideas are required.

We describe the details of our algorithm in the next section. We show an execution of our algorithm on the example in Figure 1 in Appendix A.

## 2 Algorithm

Our algorithm for NW-PCST is primal-dual, and constructs an integral feasible solution for (P) as well as a feasible solution for ($D^0$) whose values are within an $O(\ln n)$ factor of each other. This is accomplished in phases. Phase $i$ starts with a set of *initial* components $\mathcal{C}^i$ as well as a component $T_r$ that contains the root. Our algorithm maintains a tree for each of these sets, spanning its vertices. We also maintain the invariant, that no two components in $\mathcal{C}^i$ are adjacent, or in other words, no two components in $\mathcal{C}^i$ are connected by an edge.

Initially, in the very first phase of the algorithm we define these sets as follows: recall that we call a vertex cheap if its cost is at most its penalty. We look at the graph induced by the root $r$ and all the cheap vertices, i.e. $G[\{r\} \cup \{v : v \text{ is cheap}\}]$. Each connected component in this induced subgraph will be an initial component in $\mathcal{C}^1$, except for the component containing the root that will constitute $T_r$.

In phase $i$ we run a dual growing process that computes a feasible solution $(y^i, p^i)$ for ($D^0$), and either

[i] finds a tree $T^i$ connecting a set $\mathcal{C}(T^i) \subseteq \mathcal{C}^i$ of at least two initial components with cost proportional to the total value of the dual solution $(y^i, p^i)$, or

[ii] determines that the total penalty of components in $\mathcal{C}^i$ is at most the value of $(y^i, p^i)$.

In case [i] we either replace $\mathcal{C}(T^i)$ in $\mathcal{C}^i$ by the single set $T^i$, or we replace $T_r$ by $T_r \cup T^i$, while in case [ii] the algorithm terminates and returns the tree $T_r$.

To achieve the claimed approximation guarantee, we will show that at least one among the feasible dual solutions $(y^i, p^i)$ produced by the algorithm in each phase, has a value within an $O(\ln n)$ factor the cost of the returned tree $T_r$ plus the penalties of the nodes in $V \setminus T_r$.

We now give a detailed description of a phase of our algorithm. For notational convenience we will omit superscripts $i$ whenever there is no ambiguity. We will eliminate all $p$-variables, and revert to reduced costs and penalties as described in Section 1.1. Our dual growing process will now compute a dual solution feasible for (D).

The dual growing procedure of the current phase is best described as a process over time. The algorithm maintains a feasible dual solution $y^\tau$ for (D) for every *time* $\tau \geq 0$, and we let $\mathcal{S}^\tau$ be a collection of sets that contains its *support*, (i.e., $\mathcal{S}^\tau$ contains all $S$ with $y^\tau_S > 0$).

Call a vertex $v$ *tight* if the constraint (1) for $v$ holds with equality for $y^\tau$, and note that cheap vertices are always tight. In the following we will call a maximal connected set of tight vertices with respect to $y^\tau$ a *moat*. Any two moats are clearly vertex disjoint, and no two vertices in different moats can be adjacent. A moat $S$ is *active* at time $\tau \geq 0$ if

$$\sum_{U \subseteq S} y^\tau_U < \bar{\pi}(S),$$

and *inactive* otherwise. For ease of notation, we let $\mathcal{A}^\tau$ be the collection of active moats. Initially, $\mathcal{A}^0$ is the set $\mathcal{C}$ of initial components, and so each initial component is contained in an active moat at time 0. We also define $\mathcal{I}^\tau$ as the set of all inclusion-wise maximal inactive sets in $\mathcal{S}^\tau$. For $C \in \mathcal{C}$, and $\tau \geq 0$, we let
$$\text{age}^\tau(C) = \min\{\tau, \sum_{v \in C} \bar{\pi}(v)\}$$

be the *age* of $C$ at time $\tau$. The age of $C$ is the first time, where $C$ becomes part of an inactive moat during the dual growing process if that has happened before time $\tau$, and $\tau$ otherwise.

For a given set $S \in \mathcal{S}^\tau$, and an initial component $C$, we call $C$ the *core* of $S$ if $C \subseteq S$ has the largest age value among all initial components contained in $S$. We will later see that every set $S \in \mathcal{S}^\tau$ has a unique core which we will denote by $\text{core}(S)$. This allows us to extend the age-notion to sets $S \in \mathcal{S}^\tau$: we let the age of $S$ at time $\tau$ be that of its defining core.

At any time $\tau \geq 0$, the algorithm grows all sets in $\mathcal{A}^\tau$ uniformly at unit rate. Several *events* may happen during this growth process.

[A] a constraint of type (2) becomes tight for an active set $S \in \mathcal{A}^\tau$, or

[B] a constraint of type (1) becomes tight for a vertex $\tilde{v}$.

In [A], the set $S$ now becomes inactive, and moves from $\mathcal{A}^\tau$ to $\mathcal{I}^\tau$. If, after $S$ moves, $\mathcal{A}^\tau = \emptyset$, then the phase ends and we will show that condition [ii] holds. Otherwise, the growing process continues for the remaining set of moats in $\mathcal{A}^\tau$.

Consider now an event of type [B]. We say that an initial component $C \in \mathcal{C}$ *loads* $\tilde{v}$ if there is a set $S \in \mathcal{S}^\tau$ with $\tilde{v} \in \Gamma(S)$ such that $C = \text{core}(S)$. Let $\mathcal{L}^\tau(\tilde{v})$ be the set of all $C \in \mathcal{C}$ that load $\tilde{v}$. If $\tilde{v} \in \Gamma(T_r)$ or if
$$\sum_{C \in \mathcal{L}^\tau(\tilde{v})} \text{age}^\tau(C) \geq \frac{3}{2}\tau, \tag{$\star$}$$

then we will find a tree $T$ that connects $\tilde{v}$ and the cores that load it, among possibly other things. We emphasize the subtle but important departure from the familiar primal-dual theme of *active mergers*: we do not require there to be more than one active moat neighbouring $\tilde{v}$ at time $\tau$!

The tree $T$ is constructed iteratively; initially we let $T = (\{\tilde{v}\}, \emptyset)$, and we will add to it in recursive calls to two main procedures, named `FindSubTree` (`FST`) and `ConnectVertex` (`CVtx`). We will refer to $T$ as the *phase tree*.

Once the tree construction is complete, the phase ends. We remove from $\mathcal{C}$ all the initial components $C$ connected by $T$ and we either replace $T_r$ by $T_r \cup T$, in case $\tilde{v} \in \Gamma(T_r)$, or add the single component $T$ to $\mathcal{C}$. We will show that condition [i] holds for $T$.

If instead $\tilde{v} \notin \Gamma(T_r)$ and $(\star)$ does not hold, then we simply continue the growing process. Note that, by definition of moat, there will now be a single active moat containing $\tilde{v}$ as well as all the sets in $\mathcal{A}^\tau \cup \mathcal{I}^\tau$ adjacent to $\tilde{v}$. Indeed, at any time $\tau$, $\mathcal{S}^\tau$ is a laminar family of non-adjacent sets.

We provide more details for the computation of tree $T$ in [B]. The general goal of the computation is to construct a tree $T$ with the following property: for *every* expensive vertex $w \in T$, $T$ connects *all* the initial components that load $w$! To this aim, let $S_1, \ldots, S_p$ be the collection of inclusion-wise maximal sets in $\mathcal{S}^\tau$ that neighbour $\tilde{v}$. For each of these sets $S_j$ we will now invoke the function `FST` whose job is to find a tree $T_j$ that connects $\tilde{v}$ to those cores in $\mathcal{L}^\tau(\tilde{v})$ that are contained $S_j$. In general, the procedure `FST` takes as parameters two vertex sets $S \subseteq V$ and $L \subseteq \Gamma(S)$; the set $L$ contains vertices that are already spanned by the phase tree $T$. `FST` computes a tree connecting $L$ to core($S$) and adds it to $T$. In our computation of tree $T_j$, we call FST($S_j, \{\tilde{v}\}$). The final tree $T$ centered at $\tilde{v}$ will then be the union of $\{\tilde{v}\}$ and $\bigcup_{j=1}^{p} T_j$.

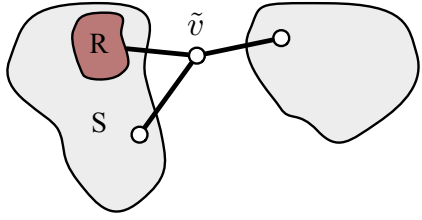

We point out the following subtlety indicated in the figure above. There may be sets $S, R \in \mathcal{A}^\tau \cup \mathcal{I}^\tau$ that both load $\tilde{v}$ and hence their cores are contained in $\mathcal{L}^\tau(\tilde{v})$. Notice however that, if $R$ is contained in $S$, then $R$ is not among the sets $S_1, \ldots, S_p$, and hence FST will not be called *directly* for $R$, but rather will be invoked *indirectly*, at some later point in the recursive procedure. This is important to ensure that FST is called *at most once* for each set in $\mathcal{S}^\tau$: indeed, this will be crucial for our analysis.

## 2.1 FindSubTree

FST($S, L$) first constructs auxiliary graph $H_S$ as follows. Let $\bar{\tau}$ be the time at which the construction of the phase tree started, and hence, where the current phase ended. Let $\tau$ be the age of $S$ at time $\bar{\tau}$. (Note that $\tau$ could be $< \bar{\tau}$, if $S$ is an inactive set at time $\bar{\tau}$). Start with graph $G[S \cup L]$ and among all the inactive sets in $\mathcal{S}^\tau$ contained in $S$, identify inclusion-wise maximal ones; abusing notation, we will refer to the super-vertex resulting from identifying such a set $R$ by $R$ as well.

Observe that $H_S$ may now have super-vertices but also original, expensive vertices. In fact, no two super-vertices are adjacent, or in other words, every neighbour of a super-vertex in $H_S$ is an original expensive vertex. Each such expensive vertex is tight (in terms of constraint (1)) at time $\bar{\tau}$. We let its *auxiliary* cost $c_S(v)$ be the total amount of dual load it feels from core($S$); i.e.,

$$c_S(v) = \sum_{\substack{R \subseteq S, v \in \Gamma(R) \\ \text{core}(R) = \text{core}(S)}} y_R^\tau.$$

Let the auxiliary costs of the super-vertices and the vertices contained in the core be 0.

FST has two main parts that we now describe.

**Part I**. The purpose of part I is to find a tree $T_L$ that connects the vertices in $L$, and to add it to $T$. We will later see that $L$ has never more than two vertices. We start by computing a minimum-$c_S$-cost path $P$ in $H_S$ connecting the vertices in $L$. This is trivial when $L$ consists of a single vertex $a$ where we let $P = \{a\}$. Otherwise, assume that $L = \{a, b\}$ for some $a, b \in \Gamma(S)$. Compute a minimum-$c_S$-cost $a, b$-path $P$; i.e., a path in $H_S$ connecting $a$ and $b$ with minimum value of

$$\sum_{v \in P} c_S(v).$$

We now add the path $P$ to the phase tree $T$; recall that the two end-points of $P$ are already in the phase tree. The path $P$ may contain super- as well as expensive original vertices. Super-vertices will be replaced by sub-trees as follows. For each super-vertex $R$ on this path with neighbours $a'$ and $b'$ we recursively call FST with parameters $R$, and $L = \{a', b'\}$. We also call

$$\texttt{FST}(R', \Gamma(R') \cap P)$$

for each super-vertex $R'$ that is not on $P$, but is the neighbour of some original expensive vertex in $P$; see Figure 2 for an illustration. The trees returned by all of these subcalls are added to the phase tree $T$, and indeed these trees together with the expensive vertices in $P$ form a tree $T_L$ connecting vertices $a$ and $b$.

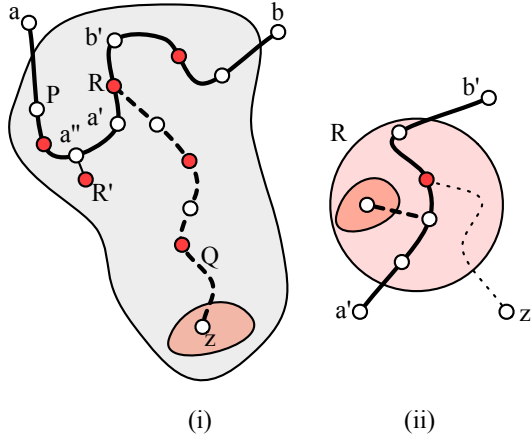

Figure 2: Inside method FST. Paths $P$ and $Q$ appear in thick solid and dashed stroke, respectively. Original (expensive) vertices appear in white, and super-vertices are coloured in red; the small red subset indicates the core of the considered set.

Why do we recursively call the function FST on the sets that neighbour expensive vertices in $P$? The reason is that their cores are loading such expensive vertices. Since those vertices will become part of the tree $T$, as outlined in the previous subsection, we now want to connect all cores that load them as well.

**Part II**. The goal in the second part of FST is to connect the core of $S$ to the phase tree $T$. This is established by procedure CVtx that takes as parameters a set $S$, a vertex $a \in \Gamma(S) \cup S$, and (for analysis purposes) a level index $i$. The procedure computes a tree contained in $S \cup a$ that connects $a$ to $T$, and adds that to the phase tree. In our specific case, we choose an arbitrary vertex $z$ in core$(S)$, and call

$$\texttt{CVtx}(S, z, 0).$$

Indeed, at the end, FST adds core$(S)$ to the phase tree as well.

## 2.2 ConnectVertex

Much like FST, CVtx$(S_d, z_d, d)$ first computes an auxiliary graph $H_{S_d}$, as follows. First, consider the graph $G[S_d \cup z_d \cup T]$, where $T$ is the current phase tree. Then, as in FST, among all sets in $\mathcal{S}^\tau$ contained in $S_d$, identify all inclusion-wise maximal inactive ones, where $\tau = \mathrm{age}^{\bar\tau}(S_d)$ and $\bar\tau$ is the time at which the current phase ended. Finally, define the auxiliary cost $c_{S_d}(v)$ of a vertex $v$ to be 0 if $v$ is a super-vertex or if $v$ is in $\mathrm{core}(S_d)$. Otherwise, let $c_{S_d}(v)$ be the total amount of dual load it feels from $\mathrm{core}(S_d)$.

The procedure computes a shortest-$c_{S_d}$-cost path $Q_d$ in $H_{S_d}$, from $z_d$ to $T$. $Q_d$ ends in some (super- or non-super-) vertex $S_{d+1}$ in $H_{S_d}$. Note that $S_{d+1}$ could be a super-vertex. However, this would mean that $S_{d+1} \cap T \neq \emptyset$, and therefore that the function FST has been called already for the set $S_{d+1}$. For this reason, we do not call the function FST again on $S_{d+1}$, but we rather recursively call

$$\mathtt{CVtx}(S_{d+1}, z_{d+1}, d+1)$$

where $z_{d+1}$ is the second-last vertex on $Q_d$. Now add the path $Q_d$ to the phase tree $T$. This path may contain super- as well as expensive vertices. For all super-vertices $S'$ on $Q_d$ we now call $\mathtt{FST}(S', \Gamma(S') \cap Q_d)$. The computed tree will replace the super-vertex placeholder in phase tree $T$.

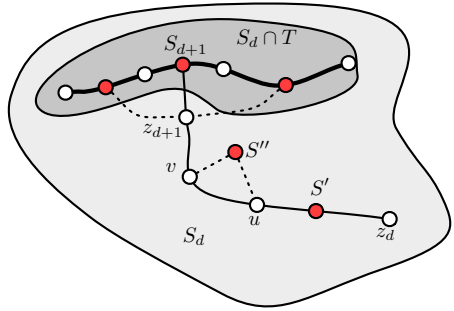

Each expensive vertex $v$ on $Q_d$ may feel dual load from (the inactive component of a) super-vertex $S''$, and hence $\mathrm{core}(S'') \in \mathcal{L}^\tau(v)$. We observe, however, that among the internal vertices, only $z_{d+1}$ may feel dual load from super-vertices in $T \cap S_d$. This follows from the fact that $Q_d$ is a shortest path according to auxiliary costs $c_{S_d}$. The figure above illustrates the situation. CVtx now calls

$$\mathtt{FST}(S'', \Gamma(S'') \cap Q_d)$$

for all super-vertices $S''$ not in $T$ that neighbour a vertex $v$ in $Q_d$.

## 3 Analysis

### 3.1 Correctness

The goal here is to show that the algorithm described in the previous section can be correctly implemented.

First of all, note that in each phase $i$ the number of initial components in $C^i$ decreases by at least one, and therefore the algorithm terminates after at most $n$ iterations.

The final output is a connected tree $T_r$, containing the root by definition. Connectivity of $T_r$ easily follows by the way FST and CVtx are defined. However, correctness of the above functions crucially relies on two properties that we just stated in the previous sections and we are now going to prove:

(a) Each set $S \in \mathcal{S}^\tau$ has a unique core;

(b) In each call of the function FST the cardinality of the set $L$ is at most 2.

To see (a), we show the following more general statement by induction: at any time $\tau$, an active set $S \in \mathcal{A}^\tau$ contains exactly one initial component $C$ with $\text{age}^\tau(C) = \tau$. This implies (a), because any set in $\mathcal{S}^\tau$ was an active set up to some time $\tau' \leq \tau$. Observe that the statement is certainly true at the beginning of each phase ($\tau = 0$). The only event that changes the family $\mathcal{A}^\tau$ at time $\tau$ is when a vertex $\tilde{v}$ becomes tight. In order for this to happen, it is necessary that at least one set $S$, with $\tilde{v} \in \Gamma(S)$, is in $\mathcal{A}^\tau$. Suppose we have two such sets $S_1$ and $S_2$. By induction hypothesis, there is one initial component in $S_1$ and one initial component in $S_2$ with age $\tau$. But in this case, condition $(\star)$ holds and the phase ends. Therefore, if the phase does not terminate, there is only one set $S \in \mathcal{A}^\tau$ with $\tilde{v} \in \Gamma(S)$. Any other maximal set $S' \in \mathcal{S}^\tau$ adjacent to $\tilde{v}$, if any, belongs to $\mathcal{I}^\tau$ and by induction hypothesis it follows that it contains only initial components with age $< \tau$. This implies that the new active set, which now includes $S, \tilde{v}$, and all its adjacent inactive sets, will contain only one initial component with maximum age value $\tau$.

To see (b), we observe that there are only three different situations that lead to FST calls. The first one is when $(\star)$ is verified: here we will call FST with $L = \{\tilde{v}\}$.

The second possibility is when there is a super-vertex $R$ that belongs to either a path $P$ computed inside a FST call, or a path $Q$ computed inside a CVtx call. In both cases, we call FST with $L = \{a, b\}$, where $a$ and $b$ are the expensive vertices on the path adjacent to $R$.

The last possibility is when there is a super-vertex $R'$ adjacent to an expensive vertex $v$ that belongs to either a path $P$ computed inside a FST call, or a path $Q$ computed inside a CVtx call. Assume $v \in P$ (the other case is identical). We call FST with $L = \{\Gamma(R') \cap P\}$. What does set $L$ look like? Recall that $P$ is a minimum-$c_S$-cost path in an auxiliary graph $H_S$ (for some set $S$). Since super-vertices in $H_S$ are never adjacent, $L$ contains only expensive vertices. We now claim that any expensive vertex $w \in H_S$ has a positive auxiliary cost $c_S(w)$. If not, it means $w$ became tight at some time $0 < \tau' < \text{age}^{\bar\tau}(S)$, where $\bar\tau$ is the time at which the current phase ended. However, when that happened we had to check condition $(\star)$: since the phase did not end, $w$ became part of an active set at time $\tau'$, and therefore it would now be inside a super-vertex, a contradiction. It follows that, if $|L| \geq 3$ then the path $P$ contains at least 3 expensive vertices with positive auxiliary cost that are adjacent to $R'$. Since the auxiliary cost of $R'$ is instead 0, $P$ could be shortcut contradicting the fact that it is of minimum cost.

## 3.2 Bounding the tree cost in phase $i$

Recall that with the exception of the very last phase, each phase $i$ of the algorithm computes a tree $T^i$ that joins a number of initial components $\mathcal{C}(T^i)$ in $\mathcal{C}^i$. The tree will then be merged with $T_r$ if it and $T_r$ have common adjacent vertices, and $T^i$ will be added to $\mathcal{C}^{i+1}$ otherwise, replacing the components of $\mathcal{C}(T^i)$. In this section we will bound the total cost of $T^i$. We do this by providing a detailed charging scheme that distributes $\bar{c}(T^i)$ over the components in $\mathcal{C}(T^i)$.

Recall from the description of phase $i$ in Section 2 that we start construction of tree $T^i$ as soon as constraint (1) becomes tight for some vertex $\tilde{v}$ and either $\tilde{v} \in \Gamma(T_r)$ or ($\star$) holds. Suppose that this happens at time $\tau \geq 0$, and recall that $y^\tau$ is the feasible dual solution for (D) at this time. By definition the set $\mathcal{S}^\tau$ contains all sets $S$ in the support of $y^\tau$, and it is not hard to see that the family of these sets is laminar (i.e., any two sets are either disjoint, or one is fully contained in the other).

As soon as a vertex $\tilde{v}$ becomes tight and either $\tilde{v} \in \Gamma(T_r)$ or ($\star$) holds, our algorithm invokes FST for each inclusion-wise maximal set $S \in \mathcal{S}^\tau$ that neighbours $\tilde{v}$; the goal being to connect the core of $S$ to $\tilde{v}$. Each of these *top-level* FST calls may itself trigger further, lower level FST calls on sets $S'$ that are descendants of $S$ in the laminar family defined by $\mathcal{S}^\tau$.

In Part I, Function FST constructs a partial tree $T_L$ to be added to $T$. In part II, FST invokes CVtx in order to connect vertex $z_0 \in \text{core}(S)$ to the partial tree $T_L$ constructed so far. In its construction of a path linking core and $T_L$, CVtx will call itself with progressively higher level-indices. A chain of CVtx calls is associated with a chain $\{S_i\}_{i=0}^p$ of sets in $\mathcal{S}^\tau$ with the following properties:

(i) $S_0 = S$,

(ii) $S_i$ is a super-vertex in $H_{S_{i-1}}$ contained in $S_{i-1}$ for all $1 \leq i \leq p-1$,

(iii) $S_p$ is an original vertex contained in $H_{S_{p-1}}$.

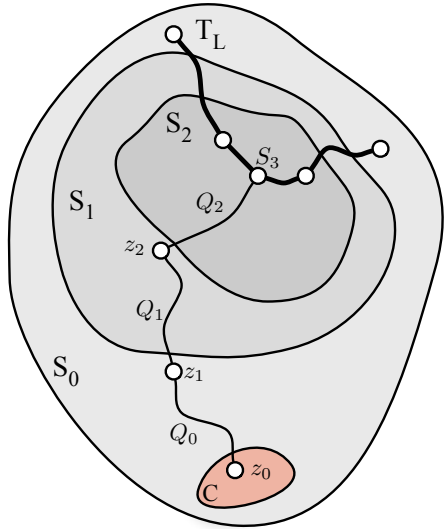

Figure 3: The chain of CVtx calls invoked from within a call to FST. The thick black lines indicate edges of the tree $T_L$ constucted by FST.

For $0 \leq i \leq p-2$, the level-$i$ call to `CVtx` computes a shortest $c_{S_i}$-cost $z_i, S_{i+1}$-path $\bar{Q}_i$ in $H_{S_i}$. By assumption $S_{i+1}$ is a super-vertex of $H_{S_i}$. Let $z_{i+1}$ be its original, expensive predecessor on path $\bar{Q}_i$, and let $Q_i$ denote its $z_i, z_{i+1}$-prefix. We then inductively call

$$\texttt{CVtx}(S_{i+1}, z_{i+1}, i+1).$$

Finally, $S_p$ is an original vertex that is part of tree $T_L$, and the $z_0, T$-path is simply the concatenation of $Q_0, \ldots, Q_{p-1}$. Figure 3 illustrates the construction for $p = 3$.

A component $C \in \mathcal{C}^i$ is charged whenever `FST` is called on set $S$ whose core is $C$. As we will see later, during a phase, `FST` is executed at most once for any given set $S$, and thus, a component $C \in \mathcal{C}^i$ is charged at most once during a phase $i$. Even stronger, we will see that any core $C$ is charged at most once throughout the *entire* execution of the algorithm. This fact will be crucial for the performance ratio analysis in the next section.

As mentioned, a call of the form `FST(S, L)` yields a *charge* $\Phi(\text{core}(S))$ to $\text{core}(S)$. This charge consists of three main components.

(C1) The auxiliary cost $c_S(P)$ of the $L$-path $P$ computed.

(C2) The auxiliary cost

$$\sum_{j=0}^{p-1} c_{S_j}(Q_j)$$

of the paths computed in the `CVtx` calls.

(C3) The non-auxiliary costs

$$\sum_{j=1}^{p-1} (\bar{c}(z_j) - c_{S_{j-1}}(z_j))$$

of vertices $z_1, \ldots, z_{p-1}$.

Several comments are in order. Consider an expensive vertex $v$ on $P \cup Q_0 \cup \ldots Q_{p-1}$. We charge the vertex' auxiliary cost to $\text{core}(S)$. Who pays for the non-auxiliary part of $v$'s cost? Recall that, in the description of `FST` and `CVtx`, if $v \neq z_1, \ldots, z_{p-1}$ we are careful to invoke `FST` for all super-vertices $S'$ whose cores are in $\mathcal{L}^\tau(v)$. The set $\mathcal{L}^\tau(v)$ contains all cores of sets in $\mathcal{S}^\tau$ that ever load $v$. Thus, through these `FST` calls, we make sure that the entire cost of all vertices in the constructed trees is charged to some cores.

Vertices $z_1, \ldots, z_{p-1}$ receive special treatment; why? A vertex $z_j$, $1 \leq j \leq p-1$ is incident to super-vertex $S_j$ in $H_{S_{j-1}}$ where $S_j$ intersects $T$. Vertex $z_j$ may also be adjacent to other super-vertices that intersect $T$. All of these super-vertices have $c_{S_{j-1}}$-cost 0. If a vertex $z_j$ feels dual load from super vertex $R$ that intersects $T$, then we may have already called `FST` on $R$. We must not call `FST` twice for a set $R \in \mathcal{S}^\tau$! For this reason, we charge all its reduced cost $\bar{c}(v)$ to $\text{core}(S)$ instead. Differently, note that $S_p$ is an expensive vertex by definition, and the vertex $z_p$ preceeding $S_p$ on $Q_{p-1}$ is not incident to super-vertices intersecting $T$. Otherwise, $Q_{p-1}$ would not be a shortest $z_{p-1}, T$-path in $H_{S_p-1}$.

We now bound the total charge $\Phi(C)$ to a core $C$ in phase $i$. Let us first show that each core is charged at most once in the entire algorithm.

**Lemma 3.1.** *Any set $C$ of vertices is charged at most once in the entire algorithm; i.e.,* `FST` *is called at most once on a set $S$ whose core is $C$.*

*Proof.* First note that whenever `FST` is called for a set $S$ with core $C$ in some phase, then $C$ is connected to at least one other core through the phase tree. Hence, $C$ will be removed from the list of initial components $\mathcal{C}$ in this phase, and will never enter it again. Hence, it suffices to show that a set $C \in \mathcal{C}$ is charged at most once in any given phase $i$.

Within phase $i$, let $S$ be the first set with core $C$ for which `FST` is called. `FST`$(S, L)$ recursively computes a tree within $G[S \cup L]$ that connects the core of $S$ (i.e. $C$) to the phase tree. All `FST` calls from within `FST`$(S, L)$ must be on inactive strict subsets of $S$, and hence all these subsets have a different core. On the other hand, after `FST`$(S, L)$ is completed, core$(S)$ is connected to the phase tree.

Finally, note that in our algorithm, we never call `FST` for sets $S$ whose core is part of the phase tree. This completes the argument. $\square$

We continue by focusing on a specific component $C \in \mathcal{C}^i$. Suppose that $C$ was charged during `FST`$(S, L)$. Recall that we let $\tau$ be the time at which the construction of the phase tree started, and hence, where the phase ended. We will bound the part of $\Phi(C)$ coming from (C1), (C2), and (C3).

**Lemma 3.2.** *The auxiliary cost $c_S(P)$ of the path connecting the vertices in $L$ is at most $2\,\mathrm{age}^\tau(S)$.*

*Proof.* By the standard primal-dual argument on shortest paths, it follows that any path from any vertex in $H_S$ to any vertex in core$(S)$ has auxiliary cost at most age$^\tau(S)$. Therefore, the shortest path between any 2 vertices in $H_S$ has auxiliary cost at most $2\,\mathrm{age}^\tau(S)$. $\square$

In order to bound (C2), we use a similar primal-dual argument, as well as $(\star)$.

**Lemma 3.3.** *Suppose that the path constructed in the* `CVtx` *call to connect $C$ to the phase tree consists of the concatenation of $Q_0, \ldots, Q_{p-1}$. The auxiliary cost $\sum_{j=0}^{p-1} c_{S_j}(Q_j)$ of this path is at most $4\mathrm{age}^\tau(S)$.*

*Proof.* Let $S_0 = S, S_1, \ldots, S_p$ be the chain of sets in $\mathcal{S}^\tau$ corresponding to the chain of `CVtx` calls invoked to construct $Q_0, \ldots, Q_{p-1}$. Using a similar argument as in Lemma 3.2 one sees that that

$$c_{S_i}(Q_i) \leq 2\mathrm{age}^\tau(S_i).$$

The lemma follows by showing that age$^\tau(S_i) < \mathrm{age}^\tau(S_{i-1})/2$ for all $1 \leq i \leq p-1$.

Observe that $S_i$ is a super-vertex in $H_{S_{i-1}}$ for all $1 \leq i \leq p-1$. Hence, there is a time $\bar\tau < \tau$ during the algorithm where the active moat containing $C$ meets $S_i$. Repeating an argument used in Section 3.1, $S_i$ could not have been active at the time as $(\star)$ would have been satisfied otherwise, and the phase would have ended. More specifically, the age of $S_i$ must have been smaller than $\bar\tau/2 < \mathrm{age}^\tau(S_{i-1})/2$. $\square$

It remains to bound the contributions in (C3).

**Lemma 3.4.** *Using the notation of Lemma 3.3, let $z_{i+1}$ be the final vertex of path $Q_i$ for all $0 \leq i \leq p-1$. The total non-auxiliary cost*

$$\sum_{i=1}^{p-1}(\bar{c}(z_i) - c_{S_{i-1}}(z_i))$$

*is at most $\mathrm{age}^\tau(S)$.*

*Proof.* This once again uses condition ($\star$). Since $S_{i+1}$ is a (super-)vertex in $H_{S_i}$ for all $0 \leq i \leq p-1$, it must be the case that at some time $\bar{\tau} < \mathrm{age}^\tau(S_i)$, an active moat containing $\mathrm{core}(S_i)$ tightens vertex $z_{i+1}$ (i.e., constraint (1) of $z_{i+1}$ becomes tight). Once again, as the phase did not end at this time, we know that the total age of components with cores in $\mathcal{L}^{\bar{\tau}}(z_{i+1}) - \mathrm{core}(S_i)$ (and hence the non-auxiliary reduced cost of the vertex) must be less than $\bar{\tau}/2$ which in turn is at most $\mathrm{age}^\tau(S_i)/2$. As $\mathrm{age}^\tau(S_i)$ is at most $\mathrm{age}^\tau(S_0)/2^i$, and hence the non-auxiliary costs of vertices $z_i$ decrease geometrically, the lemma follows. □

We obtain the following final corollary.

**Corollary 3.5.** *The total charge $\Phi(C)$ of a component $C \in \mathcal{C}^i$ is at most $7\,\mathrm{age}^\tau(C)$.*

Note that, trivially, $\mathrm{age}^\tau(C) = \sum_{S:C=\mathrm{core}(S)} y_S^\tau$. Finally note that the cost of each vertex in the phase tree $T^i$ is charged to some core in $\mathcal{C}^i$. Hence, the reduced cost of the $T^i$ is at most the sum of $\Phi(C)$ over all initial components $C \subseteq \mathcal{C}(T^i)$.

## 3.3 Approximation Factor Guarantee

In this section we prove that our algorithm is a primal-dual $O(\ln n)$ approximation algorithm for the Prize Collecting Steiner Tree problem.

We start by proving that the total reduced cost of the final tree $T$ returned by the algorithm in the last phase is within an $O(\ln n)$ factor of the value of a feasible dual solution $y^*$ to (D) computed during some phase of the algorithm.

Indeed, for the purpose of designing an LMP algorithm the following section we will prove a slightly stronger statement. The algorithm computes a feasible dual solution $y^i$ for every phase $i = 1, \ldots, m$ where $m$ is the total number of phases. For any such $y^i$, we consider the dual solution $\bar{y}^i$, obtained from $y$ by setting $\bar{y}_S := y_S$ if $\mathrm{core}(S) \subseteq T$, and $\bar{y}_S := 0$ otherwise. Clearly, $\bar{y}^i$ is a feasible dual solution.

We will prove the claimed bound by showing that it is possible to select $l = O(\ln n)$ indices $t_1, \ldots, t_l \in \{1, \ldots, m\}$ such that

$$\bar{c}(T) = O\Big(\sum_{j=1}^{l} \sum_{S} \bar{y}_S^{t_j}\Big). \tag{3}$$

From the previous section, we know that

$$\bar{c}(T) = \sum_{i:T^i \subseteq T} \bar{c}(T^i) \leq \sum_{C \in \mathcal{C}^i : C \subseteq T} \Phi(C)$$

Therefore, it will be enough for us to prove that

$$\sum_{C \in \mathcal{C}^i : C \subseteq T} \Phi(C) = O\Big(\sum_{j=1}^{l} \sum_{S} \overline{y}_S^{t_j}\Big).$$

To this aim, let us call $\overline{\mathcal{C}}^i$ is the collection of initial components in phase $i$ that have been included in the final tree $T$, that is $\overline{\mathcal{C}}^i = \{C \in \mathcal{C}^i : C \subseteq T\}$.

The proof will go as follows: first, we will define *buckets* $1, \ldots, l$ and assign every component in $\bigcup_i \overline{\mathcal{C}}^i$ to *exactly one* bucket. Secondly, we will prove that for every bucket $j$, we can identify one dual solution $\overline{y}^{t_j}$ such that the total charge of the components in bucket $j$ is within a constant factor of the value of $\overline{y}^{t_j}$.

Let us start by describing the assignment process of components to buckets.

Set $l := \lfloor \ln n \rfloor$. In order to avoid dealing with constant in the formulas later on, let us set $\Phi'(C) := \Phi(C)/7$ for every $C \in \bigcup_i \overline{\mathcal{C}}^i$ and let $\Phi'_{max}$ be the maximum $\Phi'(C)$ among all $C \in \bigcup_i \overline{\mathcal{C}}^i$. We assign a component $C$ to a bucket $1 \leq j < l$ if and only if

$$\frac{\Phi'_{max}}{2^j} < \Phi'(C) \leq \frac{\Phi'_{max}}{2^{j-1}}$$

and we assign $C$ to bucket $l$ otherwise. Clearly every component is assigned to exactly one bucket. We observe that bucket 1 is non-empty by definition, and bucket $l$ contains all small charges.

We now describe how we select one dual solution for each bucket $1 \leq j < l$. Let $t_j$ be equal to the smallest phase index $i$ such that there is some component $C \in \overline{\mathcal{C}}^i$ assigned to bucket $j$. If there are buckets with no components assigned to it, then we let $t_j$ be any arbitrary phase index. Note that for some $j$ and $j' \neq j$, we may have $t_j = t_{j'}$. We also let $t_l = t_1$, and we will show that

$$\sum_{\substack{C \text{ assigned to} \\ \text{bucket } j}} \Phi'(C) = O\Big(\sum_S \overline{y}_S^{t_j}\Big), \qquad (\diamond)$$

for all $1 \leq j \leq l$.

From the discussion above, it follows that ($\diamond$) implies the bound (3) on $\overline{c}(T)$. Therefore our goal now is to prove the above equality. We will treat the cases $j < l$ and $j = l$ separately.

Let us start by assuming $j < l$, and let $b_j := \frac{\Phi'_{max}}{2^j}$. Observe that, by construction, every component $C$ assigned to bucket $j$ satisfies: $\Phi'(C) \leq 2b_j$. Therefore, if bucket $j$ contains in total $K$ components, the left-hand side of ($\diamond$) is at most $2Kb_j$. The key idea to relate the quantity $2Kb_j$ to the value of the dual solution $y^{t_j}$ is that of introducing a *potential function* $\beta_j(i)$, defined as

$$\beta_j(i) = \sum_{C \in \overline{\mathcal{C}}^i} \min\{\overline{\pi}(C), b_j\},$$

for every bucket $j$, and for each phase $i$.

We now show how to use this potential function. Before, we just state a useful remark that follows directly from Corollary 3.5 and the definitions.

**Remark 3.6.** Let $\bar{\tau}^i$ be the time when phase $i$ terminates. If there is a component $C \in \overline{\mathcal{C}}^i$ which is assigned to bucket $j$, then $\bar{\tau}^i \geq \mathrm{age}^{\bar{\tau}^i}(C) \geq \Phi'(C) \geq b_j$. Moreover, $\bar{\pi}(C) \geq \Phi'(C)$.

The next lemma shows that the value of the potential function computed in phase $t_j$ is a lower bound on the value of the dual solution $\bar{y}^{t_j}$.

**Lemma 3.7.** $\beta_j(t_j) \leq \sum_S \bar{y}_S^{t_j}$.

*Proof.* By Remark 3.6 above, the time $\tau := \bar{\tau}^{t_j}$ when phase $t_j$ terminates is at least $b_j$. So for every component $C \in \overline{\mathcal{C}}^{t_j}$ we have that the age of $C$ at time $\tau$ is equal to $\min\{\bar{\pi}(C), \tau\} \geq \min\{\bar{\pi}(C), b_j\}$ and therefore,

$$\sum_S \bar{y}_S^{t_j} \geq \sum_{C \in \overline{\mathcal{C}}^{t_j}} \mathrm{age}^\tau(C) \geq \sum_{C \in \overline{\mathcal{C}}^{t_j}} \min\{b_j, \bar{\pi}(C)\} = \beta_j(t_j).$$

□

The following lemma is the heart of our analysis. It shows that in each phase $i$, the potential function $\beta_j$ decreases by an amount proportional to the total charge value of the components in $\overline{\mathcal{C}}^i$ assigned to bucket $j$.

**Lemma 3.8.** Consider a phase $i < m$, and let $k$ be the number of components in $\overline{\mathcal{C}}^i$ assigned to bucket $j$. Then

$$\beta_j(i) - \beta_j(i+1) \geq \frac{k}{2} b_j$$

*Proof.* Let $\tau := \bar{\tau}^i$ and $T^i$ be the subtree constructed in phase $i$. Moreover, let $\overline{\mathcal{C}}(T^i) = \{C \in \overline{\mathcal{C}}^i : C \subseteq T^i\}$ be the set of initial components of phase $i$ connected by $T^i$. If $\overline{\mathcal{C}}(T^i) = \emptyset$, then $\overline{\mathcal{C}}^i = \overline{\mathcal{C}}^{i+1}$ and $k = 0$, therefore the statement holds.

So, suppose $\overline{\mathcal{C}}(T^i) \neq \emptyset$. Note that the only difference between $\overline{\mathcal{C}}^i$ and $\overline{\mathcal{C}}^{i+1}$ is that $\overline{\mathcal{C}}^{i+1}$ does not contain the components in $\overline{\mathcal{C}}(T^i)$. Instead, either $\overline{\mathcal{C}}^{i+1}$ contains the new component $T^i$ or $T^i$ joins $T_r$. Therefore in both cases we have

$$\beta_j(i) - \beta_j(i+1) =$$
$$\sum_{C \in \overline{\mathcal{C}}^i} \min\{b_j, \bar{\pi}(C)\} - \sum_{C \in \overline{\mathcal{C}}^{i+1}} \min\{b_j, \bar{\pi}(C)\}$$
$$\geq \sum_{C \in \overline{\mathcal{C}}(T^i)} \min\{b_j, \bar{\pi}(C)\} - \min\{b_j, \bar{\pi}(T^i)\}.$$

Let the value of the right hand side of the equality above be $X$. We distinguish 3 cases depending on the value $k$ of components in $\overline{\mathcal{C}}^i$ assigned to bucket $j$.

If $k = 0$, it suffices to show that $\beta_j(i)$ is non-increasing in $i$. This is immediate as

$$X \geq \min\{b_j, \sum_{C \in \overline{\mathcal{C}}(T^i)} \bar{\pi}(C)\} - \min\{b_j, \bar{\pi}(T^i)\} = 0.$$

If $k > 1$, let $\bar{C} \in \overline{\mathcal{C}}^i$ be any component assigned to bucket $j$. Then, $\bar{\pi}(T^i) \geq \bar{\pi}(\bar{C}) \geq \Phi'(\bar{C}) \geq b_j$ by Remark 3.6, and therefore:

$$X \geq \Big( \sum_{C \in \overline{\mathcal{C}}(T^i): \Phi'(C) \geq b_j} b_j \Big) - b_j$$

$$\geq k b_j - b_j \geq (k-1) b_j \geq \frac{k}{2} b_j.$$

If $k = 1$, again $\bar{\pi}(T^i) \geq b_j$. Since $k = 1$, there is exactly one component $\bar{C} \in \overline{\mathcal{C}}^i$ merging into $T^i$ that is assigned to bucket $j$, and we have $\bar{\pi}(\bar{C}) \geq \Phi'(C) \geq b_j$. Moreover, for every $C \in \overline{C}(T^i)$ we have $\min\{\bar{\pi}(C), b_j\} \geq \min\{\mathrm{age}^\tau(C), b_j\}$ and since the termination condition $(\star)$ is satisfied, we have:

$$X \geq \big( \min\{\bar{\pi}(\bar{C}), b_j\} + \sum_{C \in \overline{C}(T^i) - \bar{C}} \min\{\bar{\pi}(C), b_j\} \big) - b_j$$

$$\geq (b_j + \sum_{C \in \overline{C}(T^i) - \bar{C}} \min\{\mathrm{age}^\tau(C), b_j\}) - b_j$$

$$\geq \min\{ \sum_{C \in \overline{C}(T^i) - \bar{C}} \mathrm{age}^\tau(C), b_j\}$$

$$= \min\{ \sum_{C \in \overline{C}(T^i)} \mathrm{age}^\tau(C) - \mathrm{age}^\tau(\bar{C}), b_j\}$$

$$\geq \min\{\frac{3}{2}\tau - \tau, b_j\} \geq \frac{1}{2} b_j \geq \frac{k}{2} b_j$$

$\square$

With the above two lemmas at hand, we are now ready to prove $(\diamond)$. Let $k_i$ be the number of the components in $\overline{\mathcal{C}}^i$ assigned to bucket $j$. For a bucket $j$ we have:

$$\sum_S \bar{y}_S^{t_j} \geq \beta_j(t_j) \geq \sum_{i=t_j}^{m-1} \frac{k_i}{2} b_j + \beta_j(m)$$

$$\geq \frac{1}{2} \sum_{\substack{C \text{ assigned to} \\ \text{bucket } j}} \Phi'(C)/2,$$

and hence

$$\sum_{\substack{C \text{ assigned to} \\ \text{bucket } j}} \Phi'(C) = O(\sum_{S \subseteq V'} \bar{y}_S^t).$$

It remains to prove $(\diamond)$ for $j = l$. At the beginning of the algorithm, we have at most $n$ initial components, and in every phase, at least two of these are merged into one common component.

Hence, the total number of distinct initial components throughout the algorithm is at most $2n$, and this is an upper bound on the total number of components that are charged in the algorithm.

We know that in phase $t_l = t_1$ there is a component $C' \in \bar{C}^{t_l}$ such that $\Phi'(C') \geq \Phi'_{max}/2$. Let $\tau := \bar{\tau}^{t_l}$. Then, $\sum_S \bar{y}_S^{t_l} \geq \text{age}^\tau(C') \geq \Phi'(C') = \Phi'_{max}/2$. Therefore

$$\sum_{C:\Phi'(C)\leq b_l} \Phi'(C) \leq 2nb_l \leq 2n\frac{\Phi'_{max}}{2^{\lfloor \ln n \rfloor}} \leq 4\Phi'_{max} \leq 8 \sum_S \bar{y}_S^{t_l},$$

as desired. Putting all together, we proved:

**Theorem 3.9.** *Let $T$ be the tree returned by the algorithm and let $y^*$ be the dual solution to (D) among $\bar{y}^1, \ldots, \bar{y}^m$ with maximum value. Then*

$$\bar{c}(T) \leq O(\ln n) \sum_S y_S^*.$$

The next lemma will complete our argument. Its proof follows trivially by the algorithm definition.

**Lemma 3.10.** *Let $y^m$ be the dual solution found in the last phase of the algorithm, then every component $S$ in the support of this solution is disjoint from $T$ and*

$$\bar{\pi}(V' \setminus T) \leq \sum_S y_S^m$$

The approximation bound now follows.

**Theorem 3.11.** *Let $OPT$ be the value (cost plus penalty) of the optimal solution for Prize Collecting Steiner Tree problem on an instance of problem with n vertices, then, Algorithm 1 finds a solution $T$ with $c(T) + \pi(V' \setminus T) = O(\ln n)OPT$.*

*Proof.* Recall that, given a feasible solution $y$ to the dual (D), setting $p_v = c(v)$ if $v$ is cheap, and $p_v = \pi(v)$ otherwise, yields a feasible solution $(y, p)$ to the dual $(D^0)$. Using weak duality together with Theorem 3.9 and Lemma 3.10, we have

$$c(T) + \pi(V' \setminus T) \leq \sum_{v \in V'} p_v + \bar{c}(T) + \bar{\pi}_r(V' \setminus T)$$

$$= O(\ln n)OPT + OPT = O(\ln n)OPT$$

$\square$

## 4 LMP Algorithm

A *Lagrangean Multiplier Preserving* (LMP) $\alpha$-approximation algorithm is an algorithm that finds a solution $F$ for an instance of the problem such that

$$c(F) + \alpha\pi(V \setminus F) \leq \alpha OPT,$$

where $OPT$ is the value of the optimal solution for the considered instance. In the following, we show how the algorithm described in the Section 2 can be used in a black-box fashion to obtain an LMP $O(\ln n)$-approximation algorithm for NW-PCST.

**Theorem 4.1.** *There is an LMP $O(\ln n)$-approximation algorithm for NW-PCST.*

*Proof.* Assuming that Algorithm 1 is an $\alpha$-approximation algorithm for NW-PCST problem, we will show how to obtain an LMP $2\alpha$-approximation.

Given an instance $(G, \pi, c)$ of NW-PCST problem, consider the following algorithm:

1. Set $\pi'(v) = 2\pi(v) - c(v)$ for every cheap vertex $v$ and $\pi'(v) = \pi(v)$ for every expensive vertex $v$;

2. Run Algorithm 1 for the instance $(G, \pi', c)$;

3. Output the tree $T$ returned by Algorithm 1.

Observe that if we denote the reduced penalties of the new instance by $\bar{\pi}'$, then $\bar{\pi}'(v) = 2\bar{\pi}(v)$ for every vertex.

By Theorem 3.9, we can find a dual solution $\bar{y}$ to (D) for the instance $(G, \pi', c)$, constructed in some phase $i < m$ of the algorithm, such that

$$\bar{c}(T) \leq \alpha \sum_{S \subseteq V': \text{core}(S) \subseteq T} \bar{y}_S. \tag{4}$$

Furthermore, if $z$ is the dual solution found in the last phase of the algorithm, then by Lemma 3.10 we have

$$\bar{\pi}'(V' \setminus T) = 2\bar{\pi}(V' \setminus T) \leq \sum_{S \subseteq V'} z_S. \tag{5}$$

**Claim.** The vector $y'$ obtained by setting $y'_S := \frac{\bar{y}_S}{2} + \frac{z_S}{2}$ for all $S \subseteq V'$, is a feasible solution to (D) for the original instance $(G, \pi, c)$.

It is easy to see that the cost constraints (1) are satisfied, since the costs (and therefore the reduced costs) never change and $y'$ is simply a convex combination of two solutions satisfying that set of constraints.

So, we only need to prove that the penalty constraints (2) are satisfied for every set $S \subseteq V'$. Let $\mathcal{C}$ be the set of the initial components considered by the algorithm at the beginning of the phase $i$ (where solution $\bar{y}$ was constructed), and let $\mathcal{D}$ be the set of initial components considered by the algorithm at the beginning of the last phase (where $z$ was constructed). Recall that each set in the support of $\bar{y}$ (resp. $z$), has a single core, and the age of the set is not greater than the total reduced penalty of its core. Moreover, by construction, the core of every set in the support of $\bar{y}$ is contained in $T$. Then, it follows that

$$\sum_{R \subseteq S}(\bar{y}_R/2 + z_R/2) = \sum_{R \subseteq S} \bar{y}_R/2 + \sum_{R \subseteq S} z_R/2$$
$$\leq \sum_{C \in \mathcal{C}: C \subseteq S \cap T} \bar{\pi}'(C)/2 + \sum_{D \in \mathcal{D}: D \subseteq S} \bar{\pi}'(D)/2$$
$$= \sum_{C \in \mathcal{C}: C \subseteq S \cap T} \bar{\pi}(C) + \sum_{D \in \mathcal{D}: D \subseteq S} \bar{\pi}(D)$$

Observe that $z$ is the dual solution constructed by algorithm during the last phase $m$, and therefore any initial component in $\mathcal{D}$ is either an initial component in $\mathcal{C}$ as well, or it is the union of a bunch of initial components in $\mathcal{C}$. Since all components in $\mathcal{D}$ became inactive, none of them intersects $T$. Therefore $\sum_{D \in \mathcal{D}: D \subseteq S} \bar{\pi}(D) = \sum_{C \in \mathcal{C}: C \subseteq S \setminus T} \bar{\pi}(D)$. We obtain

$$\sum_{R \subseteq S} y'_R \leq \sum_{C \in \mathcal{C}: C \subseteq S} \bar{\pi}(C) = \bar{\pi}(S).$$

Therefore, the penalty constraint holds for $S$ and the claim is proved.

Adding $\alpha$ times the inequality (5) to the inequality (4), we get

$$\bar{c}(T) + 2\alpha \bar{\pi}(V' \setminus T) \leq \alpha \sum_{S \subseteq V'} \bar{y}_S + \alpha \sum_{S \subseteq V'} z_S$$
$$\leq 2\alpha \Big( \sum_{S \subseteq V'} \frac{\bar{y}_S}{2} + \sum_{S \subseteq V'} \frac{z_S}{2} \Big) = 2\alpha \sum_{S \subseteq V'} y'_S,$$

and hence

$$c(T) + 2\alpha \pi(V' \setminus T) \leq$$
$$2\alpha \Big( \sum_{S \subseteq V'} y'_S + \sum_{v \in V'} p_v \Big) \leq 2\alpha OPT,$$

which means the algorithm is an LMP $2\alpha$-approximation algorithm. The last inequality follows again from the fact that setting $p_v = c(v)$ if $v$ is cheap, and $p_v = \pi(v)$ otherwise, yields a feasible solution $(y', p)$ to the dual (D$^0$). $\square$

## A    Example of Execution

We apply our algorithm to the instance shown in Figure 1.

*Phase 1.* We have $T_r = \{r\}$, while each cheap vertex $u_j$ and $w_j$, $j = 1, \ldots, n$, form an active (singleton) component $\{u_j\}$ and $\{w_j\}$, that is in $\mathcal{C}^1$. We raise the dual variables of the active components till, at time $\tau = 1$, we check condition $(\star)$ for the expensive vertex $v_1$ that got tight (indeed, since all the expensive vertices $v_j$ are tight, we could arbitrarily pick one of them). Condition $(\star)$ holds, and the tree $T^1$ we construct is simply the star centered at $v_1$. At this point the phase ends, and note that our first dual solution $y^1$ is exactly the dual solution found by the algorithm of [13].

*Phase 2.* We still have $T_r = \{r\}$, while $\mathcal{C}^2$ is now different. $\mathcal{C}^2$ contains the active component spanned by $T^1$, as well as an active (singleton) component $\{w_j\}$ for all $j \neq 1$. We raise the dual variables of the active components till, at time $\tau = 2$, the expensive vertex $x$ adjacent to the root becomes tight. The tree $T^2$ we construct is simply $T^1 \cup \{x\}$. At this point the phase ends, and note that the value of the dual solution $y^2$ is again $O(n)$.

*Phase 3.* We now have $T_r = \{r\} \cup T^2$, while $\mathcal{C}^3$ contains an active (singleton) component $\{w_j\}$ for all $j \neq 1$. We raise the dual variables of the active components till, at time $\tau = n$, they all become inactive! At this point the algorithm ends, by returning the current $T_r$. Note that the value of the solution found is $c(v_1) + c(\mathrm{x}) + \sum_{j=2}^{n} \pi(w_j) = (n+1) + 2 + n \cdot (n-1)$, and that the value of the dual solution $y^3$ is indeed $n \cdot (n-1)$!

## B    Algorithm Pseudocode

**Algorithm 1** PrizeCollectingSteiner$(G(V,E), c, \pi, r)$

$\mathcal{C}^1 \leftarrow \{S : S$ is a connected component not containing $r$ in the graph induced by cheap vertices and $r\}$
$T_r \leftarrow$ the connected component containing $r$ in the graph induced by cheap vertices and $r$.
Let $i$ denote the phase number.
**while** $\mathcal{C}^{i+1} \neq \emptyset$ **do**
    $i \leftarrow i + 1$
    Initialize $y_S^i \leftarrow 0$ for all $S \subseteq V'$, $\mathcal{A}^i \leftarrow \mathcal{C}^i$, $\mathcal{I}^i \leftarrow \emptyset$, $T^i \leftarrow \emptyset$ , $\tau^i \leftarrow 0$
    **while** $\mathcal{A}^i \neq \emptyset$ **do**
        $\epsilon_1 \leftarrow \min_{v \in V \setminus \cup_{S \in \mathcal{A}^i \cup \mathcal{I}^i} S} \{\frac{\bar{c}(v) - \sum_{S \subseteq V | v \in \Gamma(S)} y_S^i}{|\{S \in \mathcal{A}^i : v \in \Gamma(S)\}|}\}$
        $\tilde{v} \leftarrow$ the vertex minimizing the statement above
        $\epsilon_2 \leftarrow \min_{S \in \mathcal{A}^i} \{\sum_{v \in S} \bar{\pi}(v) - \sum_{R \subseteq S} y_R^i\}$
        $\epsilon \leftarrow \min\{\epsilon_1, \epsilon_2\}$
        $\tau^i \leftarrow \tau^i + \epsilon$
        $y_S^i \leftarrow y_S^i + \epsilon$ for all $S \in \mathcal{A}^i$
        $\text{age}^\tau(S) \leftarrow \text{age}^\tau(\text{core}(S)) \leftarrow \tau^i$ for all $\forall S \in \mathcal{A}^i$ and $\tau \geq \tau^i$
        **if** $\epsilon = \epsilon_2$ **then**
            **for** $S \in \mathcal{A}^i : \bar{\pi}(S) = \sum_{R \subseteq S} y_R^i$ **do**
                Remove $S$ from $\mathcal{A}^i$ and add it to $\mathcal{I}^i$
            **end for**
        **else**
            $\mathcal{N} \leftarrow \{$inclusion-wise maximal $S \in \mathcal{A}^i \cup \mathcal{I}^i : \tilde{v} \in \Gamma(S)\}$
            **if** $\sum_{C \in \mathcal{L}^{\tau^i}(\tilde{v})} \text{age}^{\tau^i}(C) < \frac{3}{2}\tau^i$ **and** $\tilde{v} \notin \Gamma(T_r)$ **then**
                Remove all sets in $\mathcal{N}$ from $\mathcal{A}^i$ and $\mathcal{I}^i$
                add $\{\tilde{v}\} \cup \left(\cup_{R \in \mathcal{N}} R\right)$ to $\mathcal{A}^i$
            **else**
                $T^i \leftarrow (\{\tilde{v}\}, \emptyset)$
                **for** $S \in \mathcal{N}$ **do**
                      $\texttt{FST}(S, \{\tilde{v}\})$
                **end for**
                **break while**
            **end if**
        **end if**
    **end while**
    **if** $T^i \neq \emptyset$ **then**
        $\mathcal{C}^{i+1} \leftarrow \mathcal{C}^i \setminus \{C \in \mathcal{C}^i : C \subseteq T^i\}$
        **if** $T^i \cap \Gamma(T) \neq \emptyset$ **then**
            Add vertices of $T^i$ to $T_r$
        **else**
            Add $T^i$ to $\mathcal{C}^{i+1}$ as a new initial component
        **end if**
    **else**
        **return** $T_r$
    **end if**
**end while**
**return** $T_r$

**Algorithm 2** FST$(S, L)$

    Construct the auxiliary graph $H_S$ and compute the auxiliary costs $c'_S$
    **if** $|L| = 1$ **then**
        $P \leftarrow L$
    **else**
        Assume that $L = \{a, b\}$ for some $a, b \in \Gamma(S)$
        $P \leftarrow$ a minimum $c'_S$ cost $a,b$-shortest path in $H_S$
    **end if**
    **for** each super vertex $R \in H_S : P \cap \Gamma(R) \neq \emptyset$ **do**
        FST$(R, P \cap \Gamma(R))$
    **end for**
    $T^i \leftarrow T^i \cup \{$ original vertices in $P\}$
    $z_0 \leftarrow$ arbitrary vertex in core$(S)$
    CVtx$(S, z_0, 0)$
    $T_i \leftarrow T_i \cup$ core$(S)$

**Algorithm 3** CVtx$(S, z, d)$

    Construct the auxiliary graph $H_S$ and auxiliary costs $c'_S$
    $\hat{T} \leftarrow T_i$.
    For every super vertex $R \in V(H_S)$ if $R \cap T_i \neq \emptyset$, identify the vertices of $R \cap \hat{T}$ to super vertex $R$ in $\hat{T}$
    $Q = q_1 \ldots q_l \leftarrow$ a minimum $c'_S$ cost $z, \hat{T}$-shortest path in $H_S$
    $T_i \leftarrow T_i \cup \{$ original vertices in $Q\}$
    **for** super vertex $R \in V(H_S) \setminus V(\hat{T}) : Q \cap \Gamma(R) \neq \emptyset$ **do**
        FST$(R, \Gamma(R) \cap Q)$
    **end for**
    **if** $q_l = R$ is a super vertex **then**
        CVtx$(R, q_{l-1}, d+1)$
    **end if**